# Quantum discreteness is an illusion [*]

**H. Dieter Zeh (www.zeh-hd.de)**
**Universität Heidelberg, Germany**

**Abstract:** I review arguments demonstrating how the concept of "particle" numbers arises in the form of equidistant energy eigenvalues of coupled harmonic oscillators representing free fields. Their quantum numbers (numbers of nodes of the wave functions) can be interpreted as occupation numbers for objects with a formal mass (defined by the field equation) and spatial wave number ("momentum") characterizing classical field modes. A superposition of different oscillator eigenstates, all consisting of $n$ modes having one node, while all others have none, defines a non-degenerate "n-particle wave function". Other discrete properties and phenomena (such as particle positions and "events") can be understood by means of the fast but continuous process of decoherence: the irreversible dislocalization of superpositions. Any wave-particle dualism thus becomes obsolete. The observation of *individual* outcomes of this decoherence process in measurements requires either a subsequent collapse of the wave function or a "branching observer" in accordance with the Schrödinger equation – both possibilities applying clearly after the decoherence process. Any probability interpretation of the wave function in terms of *local* elements of reality, such as particles or other classical concepts, would open a Pandora's box of paradoxes, as is illustrated by various misnomers that have become popular in quantum theory.

## 1. Introduction

The discreteness of Nature was the *leitmotiv* for the physics of the first third of the twentieth century. Atoms and molecules (particles, as it seemed) were confirmed to exist, while electromagnetic radiation was shown to consist of quanta (photons) that may also exhibit particle-like effects. The atoms themselves were found to be made of nuclei and electrons as the carriers of discrete units of charge, and to exist in discrete energy states formed by these constituents – with transitions between them occurring as discrete "quantum jumps". The new theory got the name "quantum theory" precisely because Max Planck had used his phenomenologically discovered fundamental constant to define energy quanta for the electromagnetic field that were able to explain the observed spectral distribution of thermal radiation.

On the other hand, the basic principle of the later formulated quantum theory, the superposition principle, always requires a continuum. Its most familiar application, the superposition of classical configurations (such as the positions of $n$ particles), gives rise to the concept of wave functions, which evolve continuously in time according to the Schrödinger equation until the system is "disturbed" by a measurement. Unfortunately, most textbooks start teaching quantum mechanics by following the early Schrödinger in extensively studying

---

single-particle problems, thus giving the impression that the wave function was kind of a spatial field (see the remark concerning the "second quantization" in Sect. 2). So it may not be surprising that many scientists still believe that there is a wave function for each electron in an atom, or that the Dirac field equation was a relativistic generalization of the Schrödinger equation. According to Schrödinger's quantization procedure or the superposition principle, wave functions are defined on *configuration* space. This leads to the generic entanglement of all systems, which had for a long time been known to be essential in atomic physics, for example, but has in general not been taken seriously as a property of physical reality.

If the classical configurations to be superposed are the amplitudes of certain *fields* rather than particle positions, the wave function becomes an entangled wave functional for all of them. However, there are also quantum systems that can *not* be obtained by quantizing a classical system – for example spinor fields or certain "intrinsic" properties, such as "color".

In this formulation of quantum theory by means of wave functions, Planck's constant is *not* used primarily to define discrete quantities ("quanta"), but rather as a scaling parameter, required to replace canonical momenta and energies by wave lengths and frequencies, respectively, – just as time is replaced by length by means of the velocity of light in the theory of relativity. I will therefore simply drop it in the following by an appropriate choice of units. The corresponding "uncertainty relations" for classical quantities can then readily be understood by the Fourier theorem (see also Sect. 5), and discrete energies by the number of nodes required for their eigenstates to fulfill certain boundary conditions. The general quantum state, which is a solution of the time-dependent Schrödinger equation, may be any superposition of these discrete states – thus recovering a continuum of possible states.

Any remaining discreteness can then enter the theory only by means of a deviation from the unitary Schrödinger evolution (such as by spontaneous "quantum jumps"), or by an added "interpretation" of the wave function – in particular in terms of independently presumed particle properties (their positions and momenta) that would spontaneously assume definite values in measurements. Stochastic decay may be the most prominent example of a quantum event. However, the measurement process (including the registration of decay events) has successfully been analyzed in terms of the Schrödinger equation by taking into account the process of environmental decoherence, which is a continuous but very fast process that mimics quantum jumps in a sense that will be discussed in Sects. 3 and 4. Decoherence leads to apparent ensembles of narrow wave packets that mimic classical points in configuration space – including those representing particles or classical fields (depending on the environmental situation).

In contrast to the Copenhagen interpretation or other ones that are based on the Heisenberg picture,[1] I will therefore here argue that a global wave function may be sufficient to describe reality. This is not meant to imply that the traditional way to connect the quantum formalism with measurement outcomes is wrong – quite the contrary, but rather that its pragmatic use of "complementary" or "uncertain" classical concepts can be avoided and instead justified in a consistent way in purely wave mechanical terms. This is in fact the basic idea underlying decoherence.[2, 3]

Many objections have been raised against such a program: the deterministic Schrödinger equation cannot *really* describe quantum jumps; one has to *presume* a particle concept in order to quantize it, and a concept of particle *numbers* to define an n-particle wave function; a wave function defined on a space of classical configurations seems to represent no more than "potentialities"; etc. I will now argue that all these objections can be overcome, and that all apparent discreteness can be deduced from a smoothly evolving global wave function (Sects. 2 and 3). This will then lead to further insights regarding some misconceptions and misnomers that are popular in quantum theory (Sects. 4 and 5).

Nothing in the following is new, except for some – perhaps radical but consistent – consequences. The purpose of this article is mainly pedagogical. My major motivation to write it up was the observation that tradition seems to be incredibly strong even in the absence of any arguments supporting it. In particular, the Heisenberg picture has led to an almost religious belief in a fundamental concept of particles because of the central role of "observables" that replace the particle *variables* of the classical formalism in quantum mechanics. Feynman graphs form another easily misleading picture, although they are no more than an intuitive tool to construct certain integrals (appearing in a perturbation series), wherein the "particle" lines represent plane waves (field modes) rather than particles.

The only unexplained discrete quantities in physics may be the electric charge and its generalizations. Although there have been attempts to understand them as topological numbers (winding numbers) – hence as a consequence of certain *classical* fields – no such explanation has as yet been confirmed according to my knowledge.

**2. Quantum numbers and "particle" numbers**
On a one-dimensional configuration space, the different solutions of the eigenvalue equation $H\psi = E\psi$ are characterized by their numbers of nodes, $n$. This Schrödinger equation is usually called "stationary", but this terminology is lent from a statistical interpretation in terms of particles, and may therefore already be misleading. According to the wave function formalism it should be called "static", since it is not based on any time dependent elements. During the early days of quantum mechanics, many physicists did not only believe in a stationary distribution of particle positions resulting from some indeterminable motion, but also that the dynamics of these quantum states consisted exclusively of quantum jumps, while the time-dependent Schrödinger equation determined no more than the corresponding probabilities.[4]

Since the d'Alembertian forming the kinetic energy operator in the Schrödinger equation measures the curvature of the wave function, and since curvature must increase with the number of nodes, the energy eigenvalues can usually be sorted according to these "quantum numbers". If one wants to avoid referring to a questionable particle picture, this operator should more correctly be called a curvature operator, as it has nothing to do with motion in wave mechanics. In the multidimensional case, the nodes are replaced by a grid of hyper-surfaces, which factorizes in an elementary way only for the familiar textbook applications that exploit symmetries.

The time-dependent Schrödinger equation then replaces energies by frequencies. However, since its generic solution is a *superposition* of many different energy eigenstates, the overwhelming occurrence of microscopic systems (unlike macroscopic ones) in these eigenstates has still to be explained – see Sect. 3.

Except for the simplest cases, exact solutions of the Schrödinger equation are difficult to find, since all variables are usually entangled in a complicated way. Only approximate or effective solutions, often in terms of phenomenological variables, are then available. Coupled harmonic oscillators provide a very fortunate exception – even when they are themselves based on an approximation. A system of coupled oscillators can be diagonalized *before* quantization, defining classical "eigenmodes" $k$, say. So these eigenmodes can be treated as independent quantum oscillators, which give rise to a product wave function – each factor with its own number of nodes $n_k$. Such a procedure applies, in particular, to free fields (continua of coupled oscillators). Here, the nodes of the wave functions have to be distinguished from the nodes that characterize the classical field modes, $k$. Wave numbers of field modes represent *spatial* rather than canonical momenta.

This situation has two important consequences: (1) energy eigenvalues of different modes add, $E = \sum E^{(k)}$, and (2) the energy eigenvalues of the individual harmonic oscillators $k$ depend linearly on $n_k$. For a Klein-Gordon field with "mass" $m$, the energy eigenvalues for the quantized field modes are $E^{(k)}{}_{n_k} = \sqrt{m^2 + k^2}\left(n_k + \frac{1}{2}\right)$. The quantum numbers $n_k$ therefore appear as "occupation numbers" for relativistic particles with mass $m$ and spatial momentum $k$. However, in this field quantization, $n_k$ is nothing but the number of nodes of the factor wave functions characterizing the modes *k*. From this point of view it is not at all surprising that there are also superpositions of different "particle numbers", in particular coherent states that may define quasi-classical fields or certain properties of Bose-Einstein condensates etc.

This interpretation of Planck's quanta as nodes of a wave function has recently been directly confirmed in an elegant experiment by determining the Wigner function for *n*-photon states in a single cavity mode.[5] The Wigner function is defined as a partial Fourier transform of the density matrix, which in the pure case is just the dyadic product of the wave function with itself. The nodes then form circles in the thereby defined formal phase space of this one-dimensional oscillator. They have nothing to do with the nodes of the "photon wave function" (in space), which is here given by the cavity mode, while the observed nodes in configuration or phase space (defining "particle numbers") varied between 0 and 4 in this experiment.

In the case of massive fields, $m \neq 0$, states with different total numbers of nodes, $n = \sum n_k$, differ by larger energies (namely by multiples of the "rest mass") than those with fixed total number. Therefore, the latter states may be more easily superposed in practice than the former ones. For *n=1*, this leads to a superposition of different modes *k*, which defines a general "single-particle" wave function (in space). For *n > 1* once-occupied modes, one obtains a superposition of products of *n* modes, corresponding to an *n*-particle wave function. Evidently, the *n* different space coordinates characterizing these *n* modes can be interchanged without changing this quantum field state, while the empirical fact that wave functions come with different exchange symmetries for boson and fermion fields, thus restricting the fermion node numbers to 0 and 1 for each mode, has been explained by using further assumptions.[6]

Therefore, this nonrelativistic approximation gives rise to the conventional *n*-particle quantum *mechanics*. In contrast, *n*-photon states are very fragile under decoherence.[5,7] It is remarkable that in this way the nodes of the occupied classical field modes become the nodes of the effective quantum mechanical wave function, while the total number of nodes of the wave functional defines the fixed "particle number". Many such "effective" quantization procedures are known in cases where other degrees of freedom are "frozen" for some reason, for example the rigid rotator or vibrational modes of various objects – leading to rotational spectra or phonons, respectively. In the relativistic case, even the quantization of particles fails; one always has to use field quantization, whereby the fields to be quantized, such as spinor fields, need not even ever appear as classical fields. Their possible amplitudes (or their loop integrals in the case of gauge fields) form the "configuration" space on which wave functionals (general superpositions) define the corresponding quantum states. We may in fact know no more than some "effective fields" as yet.

Fields that never appear classical are often confused with (and were originally discovered as) "single-particle wave functions". This has led to the misnomer of a "second quantization".[8] Many formulations of quantum field theory start by introducing so-called particle creation and annihilation operators, which can be re-interpreted as formal transition operators between wave functionals with different numbers of nodes. (Note that there are no wave functionals for the generic objects of first quantization, namely for *n*-particle wave functions instead of fields!)

In this interpretation of quantum field theory, it appears particularly paradoxical that high-energy physics is often called "particle physics". The reason is certainly that the consequences of quantum fields are mostly observed as local events, such as tracks in a bubble chamber, thus giving the perfect illusion of particles. Their superpositions have observable consequences only at low energies, in particular as bound states or Bose-Einstein condensates. At high energies, only superpositions of intrinsic properties seem to remain relevant, for example neutrino oscillations. I will now explain why particles are not even required for a probability interpretation of the wave function in the sense of Born and Pauli.

## 3. Superselection rules, localization and stochastic events

When Ernst Mach was confronted with the idea of atoms, he used to ask: "Have you seen one?" He finally accepted their existence when Loschmidt's number had been confirmed beyond doubt to be finite. At this time he could hardly imagine that this number represented no more than the number of nodes of some not-yet-known high-dimensional wave function.

However, can we today not observe individual atoms and other kinds of particles in many ways? When experimentalists store single "particles" in a cavity, this may again be understood in terms of wave functionals with one single node – but why do they often *appear* as pointlike (discrete) objects in space and time?

What we actually observe in all such cases are only "pointer positions" of appropriate measurement devices, that is, positions of macroscopic objects, such as spots on a screen, droplets in a Wilson chamber, bubbles, or clicks of a counter. So one naturally expects local or instantaneous microscopic causes for these phenomena. While particles remained an essential ingredient for Heisenberg's quantization procedure and its interpretation, Niels Bohr was more careful – at least during his later years. He presumed classical concepts only for macroscopic objects. In a recent paper,[9] Ulfbeck and Aage Bohr (Niels Bohr's son) concluded that, when the decay of a nucleus is observed with a Geiger counter, "No event takes place in the source itself as a precursor of the click in the counter …". They refer to this interpretation as "the new quantum theory", although they do not specify whether they thereby mean Niels Bohr's later interpretation or their own generalization of it. So far I agree with them, but they assume furthermore that "the wave function loses its meaning" when the click occurs.

With this latter assumption, Ulfbeck and Bohr are missing a better and more consistent description of the quantum measurement and its classical outcome. When the apparatus *and* the environment are both included into the description by a Schrödinger wave function, unavoidable interactions lead to a dislocalization[i] of the initially prepared superposition, the process known as decoherence.[2,3] This consequence may well *explain* Niels Bohr's pragmatic rules,[10] but without *postulating* any genuine classical properties to assume definite values "out of the blue",[11] and without changing the rules of logic. Although Ulfbeck and Bohr's claim may thereby even be justified in the sense that the nonlocal wave function becomes *inaccessible* to any local observer, this consequence is here derived precisely by *assuming* a global wave function that obeys the Schrödinger equation.

Since misinterpretations of decoherence (such as in terms of perturbations by, rather than entanglement with, the environment) are still quite popular, let me here briefly review its mechanism and meaning. For this purpose, assume that some effective macroscopic variable $y$ had been brought into a superposition described by the wave function $\psi(y)$. Its uncontrollable environment $\chi(z)$, where $z$ may represent very many variables, would then unavoidably and extremely fast be transformed into a state $\chi_y(z)$ that depends strongly, but uncontrollably, on $y$, while a reaction (recoil) of the macroscopic system, such as $\psi(y) \rightarrow \psi'(y)$, can often be neglected. The macroscopic superposition described by $\psi(y)$, which would represent a "Schrödinger cat", is thus (in practice irreversibly) dislocalized: it exists neither at the system any more, nor in the environment. Their combined state is entangled, as its wave function $\psi(y)\chi_y(z)$ is not a product of functions that depend separately on $y$ or $z$. In general, we do not even have an interpretation for it, since we can only perform *local* measurements. However, there do exist nonlocal microscopic states whose individual meaning is well defined, such as total angular momentum eigenstates of spatially separated objects.

So one might expect never to find any macroscopic variable in a superposition $\psi(y)$. Rather, its reduced density matrix,[ii] obtained by tracing out the environment, would be the same as that of an ensemble of narrow wave packets. This is an essential step to understand the classical appearance of the world as well as the observed quantum indeterminism. The *correspondence principle* is not only unrealistic (in assuming isolated sytems), but would also be insufficient to describe the transition from quantum to classical physics.

---

[i] The term "dislocalization" was suggested to me by John Wheeler in place of my orginal "delocalization" to characterize the transformation of local superpositions into entangled (nonlocal) ones – in contrast to objects that are merely *extended* in space.
[ii] The reduced density matrix is a useful tool in the theory of decoherence. However, it has the disadvantages of (1) depending in principle on an artificial choice of subsystems, (2) not distinguishing between reversible (virtual) and irreversible (real) decoherence, and (3) not dis-

An entangled state that includes a macroscopic variable can be created by von Neumann's unitary measurement interaction. For example, if a microscopic superposition $\phi(x)=\Sigma c_n\phi_n(x)$ is measured by means of a pointer variable *y*, this means in quantum mechanical terms

$$\Sigma c_n\phi_n(x)\psi_0(y) \rightarrow \Sigma c_n\phi_n(x)\psi_n(y) ,$$

where $\psi_n(y)$ are narrow and mutually (approximately) orthogonal wave packets centered at pointer positions $y_n$. If *y* were a microscopic variable that could be isolated from its environment, this process would represent a reversible measurement (describing "virtual" decoherence of the local superposition $\phi(x)$). But according to what has been said above, the superposition of macroscopically different pointer states $\psi_n$ is immediately and *irreversibly* decohered by its environment, giving rise to further, now uncontrollable entanglement:

$$\Sigma c_n\phi_n(x)\psi_n(y)\chi(z) \rightarrow \Sigma_n c_n\phi_n(x)\psi_n(y)\chi_y(z).$$

This superposition can never be relocalized ("recohered") any more to become accessible to a local observer. Therefore, the fast but continuous process of decoherence describes an apparent collapse into an ensemble of narrow wave packets of the pointer (that is, of quasi-classical states that may discriminate between different values of *n*).

The very concept of quantization can be understood as the conceptual reversal of this physical process of decoherence, since it formally re-introduces the superpositions that were classically missing. The effective quantum states are thus described by wave functions on the classical configuration space. It is not clear, though, whether field configurations form the *ultimate* Hilbert space basis – as it is assumed in unified field theories. String theory is just a specific playground to search for other possibilities, while there is as yet no reason to question the universal validity of the superposition principle. Insofar as Poincaré invariance remains valid, low excitations can nonetheless be classified by Wigner's irreducible representations, which may then lead to *effective* quantum fields. Their formal simplicity may even explain the mathematical beauty of emerging classical (Maxwell's or Einstein's) field equations, which has often given the impression that they must represent an ultimate truth. On the other hand, interactions between the effective quantum fields lead to their intractable entanglement, a

---

tinguishing between proper and improper mixtures (ensembles and entanglement). Unfortunately, this has led to many misunderstandings about decoherence.

situation that allows only phenomenologically justified perturbation methods in connection with appropriate renormalization procedures.

Restrictions of the superposition principle, such as those applying to macroscopic variables because of their unavoidable decoherence, are called "superselection rules". Others, for example those that exclude superpositions of different electric charges, can similarly be explained by entanglement with the environment – in this case between a local charge and the quantum state of its distant Coulomb field.[12] The Coulomb constraint, which requires this specific entanglement, can either be understood as part of the kinematics, or again as being "caused" in the form of the retarded Coulomb field of the conserved charge in its past.

The arrow of time representing the irreversibility of decoherence requires initial conditions similar to those used in classical statistical physics: all correlations (now including entanglement) must form "forks of causality" based on common local causes in their past.[13] It then follows from statistical arguments, taking into account the complexity of macroscopic systems, that these correlations usually remain irrelevant for all relevant future times. So they have no local effects, and there is no recoherence in the quantum case.

In spite of decoherence, there always remains a *global* superposition. We have two options to understand this consequence of the Schrödinger equation in order to remain in accordance with the observed world: either we assume that decoherence triggers a global collapse of the wave function by means of an unknown modification of the unitary dynamics, such that all but one of the decohered components disappear, or, according to Everett, that all unitarily arising components exist simultaneously, forming a "multiverse" that consists of many different quasi-classical worlds containing many different successors of the same observers. While the different quasi-classical "worlds" that emerge by means of decoherence of a superposition of pointer positions need not be *exactly* orthogonal (their wave functions may slightly overlap), decoherence of discrete neuronal states in the sensory system and the brain may also contribute to dynamically separate the "many minds" of an observer.[14]

Superpositions of macroscopic variables are thus permanently being decohered, for example by scattered light. While thermal radiation would suffice to cause decoherence, ordered light carries away usable and even redundant *information*. A macroscopic trajectory is then said to be "overdetermined by the future" (separately in each quasi-classical branch). This is the physical reason why the macroscopic past – in contrast to the future – appears to "already exist" and to be fixed.[13]

An *almost* classical trajectory can be observed for an α-particle in a Wilson chamber by means of repeated measurements of its position by the undercooled gas. Although its center-of-mass wave function may continuously escape from a decaying atomic nucleus in the form of a spherical wave, Mott's analysis[15] has shown how interaction of the α-particle with electrons of the gas molecules gives rise to the superposition of a continuum of narrow angular wave packets (representing quasi-rays pointing in all directions) that are correlated with ionized molecules lying along almost straight tracks. The wave function does evidently *not* lose its meaning according to Ulfbeck and Bohr when the first event on a track occurs. The ions lead to the formation of macroscopic droplets, which are in turn irreversibly decohered and documented by scattered light (a consequence not yet taken into account by Mott). Decoherence separates even branches with slightly different droplet positions along the same track. This situation differs only quantitatively from that of trajectories of macroscopic bodies by (1) not completely negligible recoil of the α-particle (leading to slight deviations from straight tracks related to quantum Brownian motion), and (2) somewhat weaker interaction of the α-particle wave functions with their environment (leading to noticeable gaps between successive "particle positions" that may be regarded as forming "discrete histories").

If recoil is strong, such as for the scattering between "particles" of similar mass in a gas, the thereby decohered variables (positions) are also localized, but they cannot follow quasi-deterministic trajectories any more. Boltzmann's stochastic collision equation is then a more realistic quasi-classical approximation than the deterministic particle mechanics from which it is traditionally derived. A particle picture for the gas molecules is thus nothing but a prejudice derived from classical physics. (See also Blood.[16] – I agree with the title of this paper, even though its author has somewhat misrepresented the concept of decoherence, and thereby missed its essential role as an objective irreversible process that leads to the *emergence* of quasi-classical properties).

Decoherence does not only explain quasi-classical states as narrow wave packets (apparent points) in the thus *emerging* "configuration" space, but also apparent decay and other *events*. If a decaying system were described by a Schrödinger equation, its time-dependence would be continuous and coherent – corresponding to a superposition of different decay times. This is known to require small deviations from an exponential decay law, which are observable under specific circumstances.[17] It demonstrates that the decay of *isolated* systems is incompatible with the assumption of stochastic decay events or quantum jumps. However, when the outgoing wave front interacts with an environment, it is decohered from the later emitted partial wave. In this way, the wave is decohered into many partial waves which correspond to different decay times, whereby the time resolution depends on the strength of the interaction. In this way, decoherence leads to an *apparent ensemble of* *stochastic events*, and to their consequence of an exact exponential decay (see Sect. 4.5 of [13]). In the case of clearly separated energy eigenvalues, as they usually exist for microscopic systems, interactions of the decay products with the environment also tend to decohere superpositions of different energies. This can be directly observed for individual atoms under permanent measurement (that is, under strong decoherence),[18,iii] and it explains why microscopic systems are preferentially found in energy eigenstates.

After decoherence has become irreversible, the global superposition can for all practical (or "operational") purposes be replaced by an ensemble. A decay or transition is then regarded as "real" rather than "virtual", even when we have not yet observed it. This may justify the pragmatic though in general insufficient interpretation of the wave function as representing "quantum information". I do not know of any discrete quantum phenomenon in space or time that can *not* be described by means of decoherence.

## 4. No spooky action at a distance

The apparent events discussed towards the end of the last section were assumed to occur in local systems. As mentioned before, they are often described instead by a collapse of the wave function. If the measured system had been entangled with another, distant system, however, the latter would then be instantaneously affected by the collapse, too.[19] An observer will not be able to recognize this before he or she is informed about both measurements and their outcomes in a whole series of correlated measurements. Nonetheless, this consequence of entanglement can *not* simply be understood as a statistical correlation between local variables[20] (hence a measurement not as a mere increase of information about them).

The concept of entanglement means that quantum *states* are generically nonlocal. On the other hand, quantum field theory is assumed to be dynamically compatible with the relativistic spacetime structure. How can such a dynamical locality ("Einstein locality") even be formulated for a kinematically nonlocal theory? If a nonlocal state changes, this change does in general not consist of local changes. Since I am here trying to argue that quantum states, represented by wave functions, may describe reality, their nonlocality must then also be regarded as real (a fundamental property of *physical* states).

Let me therefore remind you how the concept of dynamical locality enters quantum theory. In classical terms, locality would mean that there is no action at a distance: states "here" cannot instantaneously influence states "there". Relativistically, this has the con-

---

[iii] The light and dark periods, observed in these experiments while the atom is in its excited or ground state, respectively, can be understood in analogy to the apparent particle paths in a Wilson chamber (Mott's analysis applied to a two-state system).

sequence that dynamical effects can only occur within the forward light cones of their causes. Since generic (entangled) quantum states are "neither here nor there", and not even composed of states here *and* states there (quantum mechanically represented by a direct product), quantum dynamics must in general describe the dynamics of *global* states. It might thus appear to be necessarily nonlocal, too.

The concept of dynamical locality in quantum theory requires more than a formal Hilbert space structure (relativistically as well as non-relativistically). It has to presume a local Hilbert space *basis* (for example defined by particles and/or spatial fields), which spans *all* states in the form of their superpositions. Dynamical locality then means that the Hamiltonian is a sum or integral over local operators.

This framework is most successfully represented by canonical quantum field theory, characterized by the following program:

(1) Define an underlying set of fields (including a metric) on a three-dimensional (or more general) manifold.

(2) Define quantum states as wave functionals of these fields (that is, superpositions of different field amplitudes at all points of this manifold).

(3) Assume that the Hamiltonian operator H (acting on wave functionals) is defined on any simultaneity as an integral over a Hamiltonian density, written in terms of field operators at each space point.

(4) Using this Hamiltonian, postulate a time-dependent Schrödinger equation for the wave functionals.

This dynamics defines a superposition of infinitesimal local changes for the global state, which can indiviudally propagate only in accordance with the spacetime structure that is defined by the arising spacetime metric.

In effective (phenomenological) quantum field theories, dynamical locality is often formulated by means of an additional condition of "microcausality". It requires that commutators between field operators at spatially separated spacetime points vanish. This condition is partially kinematical (as it also presumes a local basis for the quantum states), and partially dynamical (as it uses the Heisenberg picture for field operators). The dynamical consistency of this microcausality condition is nontrivial, since the commutators between operators at different times should be derivable from those at equal times (that is, on an arbitrary simultaneity) by *using* the assumed Hamiltonian dynamics.

We are now ready to discuss the spacetime dynamics that describes a collapse of the wave function (von Neumann's "process 1"),

$$\sum c_n \psi_n \longrightarrow \psi_n \quad \text{with probability given by } |c_n|^2 ,$$

characterizing a measurement, for example. Its instantaneous global nature forms a major obstacle to dynamical collapse theories, which have to modify the Schrödinger equation. No *direct* evidence has ever been found for such a non-unitary modification, although a collapse is always *used* in practice – regardless of its interpretation. Let me therefore point out again that the conventional textbook presentation, which insists that the wave function is but a tool for calculating probabilities for local events, is inconsistent with many consequences of the wave function: superpositions are well known to describe individually observable ("real") properties, such as total angular momentum. These properties depend on the relative phases of all components.

The problem of an instantaneous action at a distance does *not* seem to arise in the Everett interpretation, since this is based on the assumption that the (relativistic) Schrödinger equation is universally valid and exact. This universality is also used when deriving the unavoidable process of decoherence as a dislocalization of superpositions that propagates according to relativistic causality. While decoherence explains the formation of autonomous "branches" of the wave function, *it does not explain any collapse*, since all components would stay to exist in one superposition. This means that the observer, understood as the carrier of conscious awareness, also "splits" into his physically different branch versions. So he becomes aware of *definite* measurement results in each branch, for example. This new form of a psychophysical parallelism is the essential novel (though in hindsight plausible) element of the Everett interpretation. However, this subjective transition into one *definite* (though unpredictable) component is always *taken into account* when describing the dynamics of that wave function which may represent "our" quasi-classical quantum universe. It is particularly important for the preparation of definite initial states of local systems in the laboratory. Would this subjective branching, when explicitly formulated, then not necessarily lead into the same conflict with relativity as a dynamically formulated collapse?

To understand what is going on, consider the complete process of decoherence and observation in spacetime. If the general state is represented by a wave functional *Ψ[F(r),t]* of some fundamental spatial fields *F(r)* on arbitrarily chosen simultaneities characterized by a time coordinate *t*, these states form a tensor product of local states. We can, for example, write any global state at time *t* in the form

$$\Psi = \sum_{njk} c_{njk} \, \Psi_n^{system} \, \Psi_j^{apparatus} \, \Psi_k^{environment} ,$$

and similarly for *any* choice of subsystems which are spatially disjunct, and which cover all of space. To which kind of superpositions, in which representation, and when, does the effective collapse apply? Only if we had started with an initial product state, and thereafter assumed only ideal measurement interactions (as in Sect. 3), would we have directly ended up with a *single* sum in the corresponding measurement basis.

In order to analyze the resulting decoherence as a spacetime process, we may now further subdivide the environment into arbitrary spatial subregions. For example, if "near" describes the environment within a sphere of radius defined by the distance light could have traveled since the measurement began, and "far" the environment further away, we obtain for the mentioned chain of ideal measurements

$$\Psi = ( \sum_n c_n\ \Psi_n^{system}\ \Psi_n^{apparatus}\ \Psi_n^{near}\ )\ \Psi^{far}\ ,$$

where the far-region is not yet entangled with the "system". (In general, there will be additional, here irrelevant entanglement in other variables, too.) The radius of the near-region would thereby steadily but (sub)luminally grow, while very complex processes may be going on within it. If the branching of the wave function is *defined* by this decoherence process, it does not act instantaneously, but rather like a relativistic, three-dimensional zipper. While a genuine collapse would require superluminal action, the observable correlations between space-like separated measurement results are in this Everett interpretation a consequence of the nonlocality of quantum *states*. Decoherence transforms this entanglement into apparent statistical correlations between the subsystems. However, without taking into account the role of the indeterministically splitting observer, it would not represent a resolution of the measurement problem.

There has been much dispute about *when* the (real or apparent) collapse into a definite component occurs – that is, when the measurement has been completed. We *may assume* that this is the case for a given spacetime foliation as soon as the dislocalization of a superposition has *somewhere* become irreversible (in practice, since there is no fundamental irreversibility in this unitary description). On the other hand, we do *not have to* take into account a collapse before we have become aware of the outcome. This ambiguity represents the free choice of a position for the Heisenberg cut, and it refers, strictly speaking, to each subjective observer – not even to his "friend" who may act as a mediator to tell him the result. We could *in principle* perform interference experiments with our friends. This picture is also compatible with John Wheeler's *delayed choice experiment*, where the second observer decides later in some reference frame *what* experiment he will perform, thus giving the impression of an *advanced* action at a distance (cf. the discussion of the quantum eraser in Sect. 5).

So what would this subjective observation (at the end of the measurement chain) mean in quantum dynamical description? Clearly, the "near-region" must now include this observer.

It would not suffice, though, if some of those uncontrollable (thermal) variables, which are mostly relevant for decoherence, had propagated beyond his position. It is necessary that some controllable variables, which are able to carry information (such as light), have been registered by his senses, and the message has been transferred to his consciousness – so that the latter has controllably become entangled with the variable $n$. Only then has the subjective observer split into the various branches distinguished by this measurement variable. From an objective point of view (the "bird's perspective"), no branch is ever selected.

However, such a "subjective collapse" is definitely not used in practice to define the collapsing wave function that seems to represent "our world" as a function of an appropriate time coordinate. The conventional picture identifies the global collapse with the *local* completion of the irreversible decoherence process, which gives rise to the illusion of an event. Thereafter, one may pretend that the global wave function has collapsed into *one* of its branches, although we may not yet know into which one. The initial superposition over $n$ is thus replaced by an effective ensemble that describes the lacking knowledge about the observers' subjective future according to their later participation in the branching. Since this replacement of a global superposition by an ensemble is merely a heuristic picture (not a physical process), it may well be assumed to propagate superluminally. A causal decoherence process (the above-described "zipper") starts at the location of each individual measurement – even if the measured systems happened to be entangled. Their outcomes are in general macroscopically documented in some way, and therefore appear as part of a fixed, objective history to all later observers in each resulting branch. The observers' *passive* participation in the branching when getting entangled with the measurement outcomes appears to them as a "mere increase of their knowledge", thus justifying the conventional textbook description. If measurement results did instead enter existence in local stochastic events, these events would have to influence each other in a spooky way in order to explain Bell type correlations.[19]

The prejudice that reality must consist of local events is particularly persistent. The usual Copenhagen pragmatism may therefore be characterized by the position: "Better no reality at all than a nonlocal one." However, denying the reality of the wave function, and nonetheless regarding it as a carrier of "quantum information" comes pretty close to arguments used in esoteric circles.

## 5. Other related misconceptions and misnomers in quantum theory

This description of a branching wave function as a causal spacetime process demonstrates that the infamous "spooky action at a distance" is a misconception. The observed statistical correl-ations between pairs of measurement results are a consequence of entanglement, that is, of a nonlocal reality. They can be confirmed by an observer only when information from both measurement outcomes has arrived at his position. In the case of a Bell state formed by two spinors, for example, and parallel analyzers in the two spin measurements, there are only two rather than four final Everett branches in spite of two branchings caused independently at different places.

There are many experiments demonstrating quantum "weirdness". They appear as paradoxes only if one subscribes to the folklore that reality consists of local events (which have to occur spontaneously and "outside the laws of Nature" according to Pauli). These weird phenomena were in fact all predicted by *consistently using* the nonlocal wave function. Similarly, all those much-discussed no-go theorems apply to presumed local variables or their "values". The only remaining (but indeed very deep) quantum weirdness is the kinematical nonlocality: we seem to live in a high-dimensional space that appears to us as a classical configuration space only because of unavoidable decoherence.

Let me discuss a few related examples of quantum weirdness that have recently become popular. In particular, *quantum teleportation* has been celebrated as one of the most sensational discoveries in quantum theory.[21] The teleportation protocol for the usually considered spin or polarization states consists of three steps:

1. the preparation of an appropriate Bell state of two spinors by Alice and Bob, who then travel to different places – each of them keeping one of the two entangled spinors,

2. the later measurement of another (local) Bell state by Alice, involving her and a third spinor, with Alice then sending a message about the outcome to Bob, and

3. a unitary transformation performed locally by Bob on the spinor he kept with him.

It is evidently the crucial last step that has to reproduce the (possibly unknown) third spinor state, which must be destroyed by the measurement at Alice's place, at Bob's place. The first two steps are only required to inform Bob about what to do among a small set of formal possibilities without knowing the precise quantum state that is to be reproduced.

This may be more dramatically illustrated by means of a *complex* physical state to be "teleported", such as Captain Kirk (*CK*), instead of a spinor. According to the protocol, Bob would then need a device that allows him to physically transform any superposition of (a specific quantum state of) *CK* and some state in which he is absent, $a|CK> + b|NoCK>$ (as a new version of Schrödinger's cat), into any other such superposition – including a transformation of "*no CK*" ($a = 0$) into the state representing Captain Kirk ($b=0$)! This macroscopically unrealistic, though in quantum theory formally conceivable device (that in this case would even have to violate conservation rules) would evidently have to contain all the information about *CK*'s physical state. Bob would thus have to be able to locally *reconstruct* Captain Kirk when physically realizing the unitary transformation, while the first two steps of the protocol only serve to circumvent the no-cloning theorem in the case of an unknown initial Schrödinger cat superposition at Alice's place if it is this that is to be teleported.

Even for this first part of the protocol, no teleportation is actually involved. Before traveling to their final positions by ordinary means, Alice and Bob have to prepare an appropriate Bell state (for spinors or Captain Kirk occupation number states 0 and 1), and then take their now entangled subsystems with them, thereby carefully shielding them against the environment in order to avoid decoherence (impossible in practice for a macroscopic system). This nonlocal Bell state has several non-vanishing components which factorize in such a way that the subsystem carried by Bob is in one of the states that he is later supposed to unitarily transform into the required one. So all these states (and the information they represent) are physically at Bob's place before the "teleportation" proper begins. Decoherence between the four different possible outcomes of Alice's subsequent local Bell state measurement, required by the protocol, leads to four Everett branches – corresponding to the four possible Bell states. They are correlated with Bob's subsystem through the entanglement of the initial (nonlocal) Bell state. Bob himself gets correlated with Alice's measurement result when he receives her message. Therefore, he can perform different unitary transformations in the four different branches, all leading to the same intended final spinor or *CK* state – and this is called quantum teleportation.[22] (See also E. Joos on p. 173 of Ref. 3.)

Any interpretation of this experiment in local kinematical terms would have to assume some spooky action at a distance or telekinesis that has to *create* a certain local state at Bob's place, where he has then to apply his specific unitary transformation. For an interpretation in terms of nonlocal states one has instead to conclude that the quantum teleportation protocol allows one neither to teleport physical objects, *nor the information* needed to reconstruct them (even by technically unrestricted means). This object or state must be prepared in advance at its later destination as a component of an entangled state! If, in particular, the unknown state to be teleported happens to be CK, there would have to be two CKs at different places from the beginning!

Another recently invented drastic misnomer in quantum physics is the *quantum eraser*,[23] since this name seems to imply that the essential element of this procedure to recover coherence between different "possible" results of an intervening (for example "which way?") measurement was a mere destruction of the information about the latter's outcome. However, the physical destruction of information (for example, by its deterministic transformation into heat – as in the "reset" of a memory device[24]) would cause further decoherence rather than recoherence. Decoherence is precisely defined as such an irreversible transformation of a controllable superposition into uncontrollable entanglement with the environment. The generic decoherence-producing environment can thus *not* be regarded as an informed "witness": any information about phase relations is effectively erased (made useless) by decoherence. In contrast, there usually exists redundant information about quasi-classical macroscopic quantities – mainly in the form of scattered light – that gives rise to a documented macroscopic "history".[13] Only in the case of a *virtual* (reversible) measurement of a microscopic system by another one, where decoherence may be avoided for some time, can the *conjugate* variable still be measured after exactly reversing the virtual measurement and *all* its consequences (rather than merely erasing the information).

This concept of virtual measurements (including the idea of a delayed choice for the subsequent *real* measurement) was first discussed in a quantum-optical setting by Edward Jaynes.[25] Later versions of the experiment added other, more practicable reversible elements to be used as "virtual detectors".[26] While a Heisenberg cut for applying the probability interpretation has in the Copenhagen interpretation always to be chosen *ad hoc* (though far enough to keep all controllable entanglement on its quantum side), a natural boundary between quantum and quasi-classical descriptions is defined by the first irreversible occurrence of decoherence in a measurement chain (cf. Sect. 4).

Other quantum misnomers based on an inappropriate application of classical concepts have become established tradition. Examples are the *uncertainty relations* or the concept of *quantum fluctuations*. If a quantum state (state vector) is completely defined (pure), it is *certain*, and can even be confirmed by an appropriate measurement. The "uncertainty relations" apply to conjugate Fourier variables that may be used to represent the wave function – just as for classical radio waves. Nonetheless, uncertain initial conditions for non-existent classical variables have occasionally been made responsible for the observed dynamical quantum indeterminism, although an indeterminism of quantum states would instead require a (genuine or effective) stochastic modification of the Schrödinger equation.

Various kinds of "quantum fluctuations" (in particular *vacuum fluctuations*, often visualized in terms of "virtual particles") are used to describe genuine quantum properties, such as the minimum curvature of the wave function or some entanglement that exists in the static ground states of interacting quantum fields (their physical vacua). In the relativistic case, partial volumes (subsystems of the field) are described by "mixed" reduced density matrices, while the assumed virtual particles are said to become "real" as soon as their entanglement leads to macroscopic consequences by means of an irreversible decoherence process.[27] Conversely, states of relativistic quantum fields that are physically restricted to a finite volume cannot correspond to definite particle numbers if "particles" are defined in terms of (infinite) plane wave modes of a Klein-Gordon or Dirac field.[28] As a consequence, a uniformly accelerated detector, which would relativistically define a spacetime horizon as a boundary, must experience the inertial vacuum as a temperature bath – a phenomenon related to Hawking radiation. This demonstrates that the pragmatic particle concept is not only restricted to free fields – it would furthermore depend on the choice of the reference frame. However, all these phenomena are compatible with the existence of some *real* (observer-independent) global quantum state.

A thermal equilibrium is quantum mechanically described by a density matrix that represents a canonical ensemble of energy eigenstates (that is, it does not refer to any thermal motion). A classical thermal equilibrium is instead described by an ensemble that has to be justified by *time-averaging* over some chaotic motion (using ergodic theory). These ensembles of microscopic states are often regarded as "thermodynamic states". This terminology and its classical interpretation may be partly responsible for the misleading picture of "quantum fluctuations" to characterize *static* entangled quantum states.

So it seems that paradoxes and conceptual inconsistencies (often replaced by new words,[iv] such as dualism, complementarity, quantum information, etc.) arise only when one insists on classical or other local descriptions of physical reality. The resulting "weird" phenomena – including the violation of Bell's inequality – were in fact all predicted by means of the nonlocal wave function or state vector. Therefore, they could readily have been discussed as mere *gedanken* experiments by *assuming* the quantum theory to be universally valid. This can even be done for experiments that can hardly ever be performed in practice (for example, interference experiments with conscious observers). In view of the many sophisticated experiments that have already been done, though, the traditional lame excuses "The wave function describes only information" or "Quantum theory is not made for macro-scopic objects" are neither helpful nor convincing any more. In the quantum formalism, there always exists a formal though nonlocal observable (namely, the projector onto the actual state) that would give the value one with certainty, so that the unitary evolution of the state vector of a closed system could *in principle* always be traced and confirmed.

---

[iv] Mephisto: "Denn eben wo Begriffe fehlen, da stellt ein Wort zur rechten Zeit sich ein." in: *Faust I*, J.W. von Goethe.

**Acknowledgment**: I wish to thank Erich Joos for his critical comments on the manuscript.

**References**

[1] P. Mittelstaedt, *The Interpretation of Quantum Mechanics and the Measurement Process* (Cambridge University Press, 1998).

[2] H.D. Zeh, "On the interpretation of measurement in quantum theory", Found. Phys. **1**, 69 (1970); W.H. Zurek, "Pointer basis of quantum apparatus: Into which mixture does the wave packet collapse?", Phys. Rev. **D24**, 1516 (1981); E. Joos and H.D. Zeh, "The emergence of classical properties through interaction with the environment", Z. Phys. **B59**, 223 (1985); M. Schlosshauer, *Decoherence and the quantum-to-classical transition* (Springer, Berlin, 2007).

[3] E. Joos, H.D. Zeh, C. Kiefer, D. Giulini, J. Kupsch, and I.-O. Stamatescu, *Decoherence and the Appearance of a Classical World in Quantum Theory* (Springer, Berlin, 2003).

[4] H.D. Zeh, "Time in Quantum Theory", in *Compendium of Quantum Physics*, D. Greenberger, K. Hentschel, and F. Weinert, edts. (Springer, 2009) – http://arxiv.org/abs/0705.4638 .

[5] S. Deléglise, I. Dotsenko, C. Sayrin, J. Bernu, M. Brune, J.-M. Raimond, and S. Haroche, "Reconstruction of non-classical cavity fields with snapshots of their decoherence", Nature 455, 510 (2008). Cf. also Fig. 3.14 of [3].

[6] J.M. Leinaas and J. Myrheim, "On the theory of identical particles", Nuovo Cim. **B37**, 1 (1977).

[7] O. Kübler and H.D. Zeh, "Dynamics of quantum correlations", Ann. Phys. (N.Y.), **76**, 405 (1973).

[8] H.D. Zeh, "There is no 'first' quantization", Phys. Lett. **A309**, 329 (2003) – http://arxiv.org/abs/quant-ph/0210098 .

[9] O. Ulfbeck and A. Bohr, "Genuine Fortuitousness: Where did that click come from?", Found. Phys. **31**, 757 (2001).

[10] M. Schlosshauer and C. Camilleri, "The quantum-to-classical transition: Bohr's doctrine of classical concepts, emergent classicality, and decoherence", http://arxiv.org/abs/0804.1609.

[11] H.D. Zeh, "There are no quantum jumps, nor are there particles", Phys. Lett. **A172**, 189 (1993).

[12] D. Giulini, C. Kiefer, and H.D. Zeh, "Symmetries, superselection rules, and decoherence", Phys. Lett. **A199**, 291 (1995).

[13] H.D. Zeh, *The physical basis of the direction of time*, 5$^{th}$ edn. (Springer, Berlin, 2007) – see www.time-direction.de .

[14] M. Tegmark, "Importance of quantum decoherence in brain processes". Phys. Rev. **E61**, 4194 (2000) – http://arxiv.org/abs/quant-ph/9907009 .

[15] N.F. Mott, "The wave mechanics of α-particle tracks", Proc. R. Soc. London, **A126**, 79 (1929).

[16] C. Blood, "No evidence for particles", http://arxiv.org/abs/0807.3930v1.

[17] S.R. Wilkinson, C.F. Barucha, M.C. Fischer, K.W. Madison, P.R. Morrow, Q. Niu, B. Sundaram, and M.G. Raizen, "Experimental evidence for non-exponential decay in quantum tunnelling", Nature, **387**, 575 (1997).

[18] Th. Sauter, W. Neuhauser, R. Blatt, and P.E. Toschek, "Observation of quantum jumps", Phys. Rev. Lett. **57**, 1696 (1986).

[19] R. Garisto, "What is the speed of quantum information?" http://arxiv.org/abs/quant-ph/0212078 ; D. Salart, A. Baas, C. Branciard, N. Gisin, and H. Zbinden, "Testing spooky action at a distance", Nature **454**, 861 (2008) – http://arxiv.org/abs/0808.3316 .

[20] J.S. Bell, "On the Einstein-Podolsky-Rosen paradox", Physics **1**, 195 (1964).

[21] C.H. Bennett, G. Brassard, C. Crépau, R. Jozsa, A. Peres, and W.K. Wootters, "Teleporting an unknown quantum state via dual classical and Einstein-Podolsky-Rosen channels", Phys. Rev. Lett. **70**, 1895 (1993).

[22] C. Timpson, "The grammar of 'teleportation' ". Brit. J. Phil. Sci. **57**, 587 (2006) – http://arxiv.org/abs/quant-ph/0509048 .

[23] M.O. Scully and K. Drühl, "Quantum eraser: A proposed photon correlation experiment concerning observation and 'delayed choice' ", Phys. Rev. **A25**, 2208 (1982).

[24] C.H. Bennett, "Demons, Engines, and the Second Law", Sci. Amer. **257**(5), 88 (1987).

[25] Jaynes in: *Foundation of Radiation Theory and Quantum Electronics*, A. Barut, edt., Plenum 1980; see also F. Herbut, "On EPR-type entanglement in the experiments of Scully *et al*. I. The micromaser case and delayed choice quantum erasure", http://arxiv.org/abs/0808.3176v1.

[26] See Sect. 20.3 of M.O. Scully and M.S. Zubairy, *Quantum Optics* (Cambridge UP 1997).

[27] C. Kiefer, *Quantum Gravity*, 2$^{nd}$ edn. (Oxford Science Publications 2007), p.310 ff.; C. Kiefer, I. Lohmar, D. Polarski, and A.A. Starobinski, "Pointer states for the primordial fluctuations in inflationary cosmology", Class. Quantum Grav. **24**, 1699 (2007).

[28] N.D. Birrell and P.C.W. Davies, *Quantum fields in curved space* (Cambridge UP 1982).